\documentclass[letterpaper, 10 pt, conference]{ieeeconf}
\IEEEoverridecommandlockouts
\usepackage{amsmath,amssymb,amsfonts}
\usepackage{graphicx}
\usepackage{textcomp}
\usepackage{xcolor}
\usepackage{float}
\usepackage{bm}
\usepackage[hidelinks]{hyperref}
\usepackage[capitalize]{cleveref}
\usepackage{url}
\usepackage{subcaption}
\usepackage{diagbox}
\usepackage{algpseudocode}
\usepackage{algorithm}
\usepackage{hhline}
\usepackage{siunitx}
\def\BibTeX{{\rm B\kern-.05em{\sc i\kern-.025em b}\kern-.08em
    T\kern-.1667em\lower.7ex\hbox{E}\kern-.125emX}}

\crefname{paragraph}{paragraph}{paragraphs}
\Crefname{paragraph}{Paragraph}{Paragraphs}
\title{\LARGE \bf A Hybrid Framework for Efficient Koopman Operator Learning }

\author{Alexander Estornell*, Leonard Jung*, Alenna Spiro*, Mario Sznaier, Michael Everett%
\thanks{*Authors contributed equally. All authors are with Northeastern University, Boston, MA, USA. e-mail: {\tt \small \{estornell.a,jung.le,\allowbreak spiro.a,m.everett\}@northeastern.edu}, {\tt \small msznaier\allowbreak@ece.northeastern.edu}}}

\begin{document}

\maketitle

\begin{abstract}
     Koopman analysis of a general dynamics system provides a linear Koopman operator and an embedded eigenfunction space, enabling the application of standard techniques from linear analysis. However, in practice, deriving exact operators and mappings for the observable space is intractable, and deriving an approximation or expressive subset of these functions is challenging. Programmatic methods often rely on system-specific parameters and may scale poorly in both time and space, while learning-based approaches depend heavily on difficult-to-know hyperparameters, such as the dimension of the observable space. To address the limitations of both methods, we propose a hybrid framework that uses semidefinite programming to find a representation of the linear operator, then learns an approximate mapping into and out of the space that the operator propagates. This approach enables efficient learning of the operator and explicit mappings while reducing the need for specifying the unknown structure ahead of time. \\
     \textbf{Code: }\url{https://github.com/neu-autonomy/HybridKoopman}

\end{abstract}

\section{Introduction}
Within the study of estimation and controls, we often aim to understand or forecast the future behavior of a given system. For linear systems, characterizing attractive or divergent behavior is well understood. In contrast, characterizing the future of trajectories generated by nonlinear systems is difficult due to inherent transient variability. Koopman Theory has emerged as a common perspective in modeling such nonlinear systems \cite{doi:10.1073/pnas.17.5.315}, where the states of a nonlinear system are mapped to an infinite-dimensional space of observables equipped with a linear Koopman operator, allowing for analysis from established linear systems theory.

In particular, the leading eigenfunction, eigenvalues, and modes of the observable space produce a system in a space that appears linear. Early numerical methods for determining these spectral properties for linear observables broadly fall under the branch of Dynamic Mode Decomposition \cite{SCHMID_2010,ROWLEY_MEZIĆ_BAGHERI_SCHLATTER_HENNINGSON_2009,jose_nathan_kutz_brunton_brunton_proctor_2017}. To capture nonlinear dynamics, extended DMD (EDMD) proposes finding a nonlinear mapping and inverse from the state space to the linear observable space \cite{Williams_2015}, although finding the mapping is challenging. Thus, learning-based approaches \cite{Lusch_2018, doi:10.1073/pnas.1906995116, 8815339,doi:10.1137/18M1177846, potter2022dynamics} use the computational strength of neural networks (NNs) to find the Koopman operator and embeddings into the observable space through training on large amounts of data. In particular, these methods often use the latent space of an autoencoder (AE) as the observable space, meaning that the dimensionality of the observable space becomes a hyperparameter that must be chosen a priori. The dimension of this observable space is an important hyperparameter. Specifically, if chosen too small, the observable space would struggle to capture the most important eigenvalues of the system. If chosen too large, the resulting model may be too computationally expensive to train. These methods also often require prescribing the order of the underlying dynamics, which becomes another important hyperparameter --- especially when considering more complicated or chaotic systems.

In contrast to learning approaches, the method proposed in \cite{sznaier2021convexoptimizationapproachlearning} frames finding the mapping as an optimization problem and performs convex relaxations to solve for the observable space embedding via a semi-definite program (SDP). Unfortunately, that approach scales quadratically both in time (with data when solving the SDP) and in space (when embedding the states). These factors limit its applications in real-time use. 

To bridge the gap between these two approaches and to leverage their unique advantages while alleviating their drawbacks, this paper proposes a hybrid approach to finding embeddings that yield linear representations of nonlinear dynamics. To do this, we develop a two-stage optimization scheme which first uses an SDP to find an approximate observable space and Koopman operator, then applies this knowledge to train an autoencoder to find the mappings into the observable space and refine the operator. As shown in our experiments across several systems, including the chaotic Lorenz oscillator in \cref{experiments}, exploiting the information provided by the SDP to design and train the AE leads to a substantial reduction in approximation error and computation time compared a baseline AE approach \cite{Lusch_2018}.

\section{Background}
In this section, we provide a brief overview on Koopman theory, autoencoder-based methods, and convex optimization approaches for finding a Koopman operator and observables. 

\subsection{Koopman Operator Theory}
This work focuses on discrete-time dynamical systems of the form:
\begin{equation}
    \bm{x}_{i+1} = f(\bm{x}_i),
\end{equation}
where $\bm{x}_i \in \mathbb{R}^n$ and $f$ is a non-linear function. 

For any deterministic, discrete-time dynamical system, there exists a Koompman operator, $\mathcal{K}$, that is able to propagate observable functions forward in time \cite{BEVANDA2021197}:
\begin{equation}
    \mathcal{K} \circ\psi(\bm{x}_k)=\psi(\bm{x}_{k+1}),
\end{equation}
where $\psi$ is some $m$-dimensional measurement of the state, $\bm{x}_k$. Since $\mathcal{K}$ operates on the Hilbert space of observable functions, it is infinite dimensional. In particular, many methods aim to find eigenfunctions of the Koopman operator,
\begin{equation}
    \mathcal{K}\psi(\bm{x}_i)=\lambda \psi(\bm{x}_k),
\end{equation}
resulting in a linear Koopman operator that can be propagated with the corresponding eigenvalue, $\lambda$. In addition to linearity, this set of eigenfunctions $[\psi_1, \dots,\psi_n]$ spans an invariant subspace (i.e., applying the Koopman operator to any element of this subspace remains in the subspace). By finding a finite set of eigenfunctions and their corresponding eigenvalues, a finite matrix Koopman can approximate the overall dynamics of the system. Thus, the dimensionality of the embedding $m$ is an important choice when finding the Koopman operator.  

\subsubsection{Koopman Operator for Chaotic Systems} \label{time-delay-hankel}
For chaotic systems, where there may be unpredictable switching behavior or multiple fixed points, finding an accurate linear embedding is difficult; in fact, for the latter case, a continuous one-to-one linear immersion does not exist \cite{LIU202360}. A common technique is to supplement the current measurement with previous history as \emph{memory}; works such as HAVOK \cite{HAVOK17} utilized the link between the time-delay Hankel matrix and Koopman operator
\begin{align}
\left[\begin{array}{cccc}
\bm{x}_{0} & \bm{x}_{1} & \dots & \bm{x}_{q-1}\\
\bm{x}_{1} & \bm{x}_{2} & \dots & \bm{x}_{q}\\
\vdots & \vdots & \ddots & \vdots\\
\bm{x}_{r-1} & \bm{x}_{r} & \dots & \bm{x}_{r+q-2}
\end{array}\right]	= \\
\left[\begin{array}{cccc}
\bm{x}_{0} & \mathcal{K}\bm{x}_{0} & \dots & \mathcal{K}^{q-1}\bm{x}_{0}\\
\mathcal{K}\bm{x}_{0} & \mathcal{K}^{2}\bm{x}_{0} & \dots & \mathcal{K}^{q}\bm{x}_{0}\\
\vdots & \vdots & \ddots & \vdots\\
\mathcal{K}^{r-1}\bm{x}_{0} & \mathcal{K}^{r}\bm{x}_{0} & \dots & \mathcal{K}^{q+r-2}\bm{x}_{0},
\end{array}\right]	
\end{align}
to find a Koopman representation through a singular value decomposition (SVD) of the Hankel matrix. Although this method successfully approximates nonlinear chaotic systems, selecting the order of the linear system and memory size is critical for accurate recreation; selecting order too high results in large models, while too small order may result in poor prediction. 

\subsection{Finding linear embeddings through deep learning}
Leveraging neural networks' ability to learn from large datasets, \cite{Lusch_2018} proposes to use an autoencoder to learn both the Koopman operator and an embedding onto its eigenfunctions (observables).
The objective is for the encoder, $\varphi(\cdot)$, to map states, $\bm{x}\in\mathbb{R}^n$, to coordinates, $\bm{y}=\varphi(\bm{x})\in\mathbb{R}^m$, that span the eigenfunctions of the learned Koopman operator, $\hat{\mathcal{K}}^{\text{NN}}$.

The intrinsic coordinates are next passed through an auxiliary network to predict the Koopman operator's $m$ eigenvalues. A key insight of \cite{Lusch_2018} is that by allowing the eigenvalues to vary, their architecture can succinctly describe the dynamics of both systems with discrete or continuous spectrum. The network's ability to learn the Koopman operator and these intrinsic coordinates is due to three loss functions that we outline next.

\paragraph{Reconstruction Loss}
The encoder component learns an embedding into the observable space $\bm{y} = \varphi(\bm{x})$, while the decoder learns the inverse back into the state space $\bm{x} = \varphi^{-1}(\bm{y})$. Reconstruction accuracy of the system is defined by the following loss:
\begin{equation} \label{eq:reconstruct_loss}
    L_{\text{reconstruct}}(\bm{x}) = ||\bm{x} - \varphi^{-1}(\varphi(\bm{x}))||_2^2.
\end{equation}

\paragraph{Linearity Loss}
To preserve the linear dynamics of the system, a single step forward prediction can be described as $\bm{x}_{k+1} = \varphi^{-1}(\hat{\mathcal{K}} \varphi(\bm{x}_{k}))$. More generally, linearity can be enforced over $t$ time steps with the following loss:
\begin{equation} \label{eq:lin_loss}
    L_{\text{linearity}}(\bm{x}_k, t, \hat{\mathcal{K}}^\text{NN}) = ||\varphi(\bm{x}_{k+t}) - (\hat{\mathcal{K}}^\text{NN})^t\varphi(\bm{x}_k)||_2^2.
\end{equation}

\paragraph{Forward Loss}
The final property the network must learn is that the intrinsic coordinates must enable future state prediction, which is captured in the following loss:
\begin{equation} \label{eq:for_loss}
    L_{\text{forward}}(\bm{x}_k, t, \hat{\mathcal{K}}^\text{NN}) = ||\bm{x}_{k+t} - \varphi^{-1} \Bigl((\hat{\mathcal{K}}^\text{NN})^t\varphi(\bm{x}_k) \Bigr)||_2^2.
\end{equation}

\subsection{Approximating the Koopman using an SDP}
As an alternative to AE-based methods, Equations (10)-(12) of \cite{sznaier2021convexoptimizationapproachlearning} describe how to produce observables of the input set and a Koopman operator with a rank minimization SDP. In particular, that method:
\begin{enumerate}
\item Finds a dictionary of observable functions stored within a Loewner matrix, \cite{ionita2013lagrange},
\item Determines the Koopman operator, observable embedding dimension, and order of the dynamics, and
\item Solves a convex relaxation of a rank-minimization SDP, with a certificate on the optimality of the original (NP-hard) SDP.
\end{enumerate}

However, even solving the relaxed SDP can be computationally expensive, scaling in computational time with the number of data points. This limits its applicability to large datasets. Additionally, the resulting dictionary of observable functions must be decoded from a Loewner matrix, which also scales quadratically in size with the number of data points, further limiting its use with large datasets.

\section{Methodology}
\begin{figure*}
    \centering
    \includegraphics[width=0.98\linewidth]{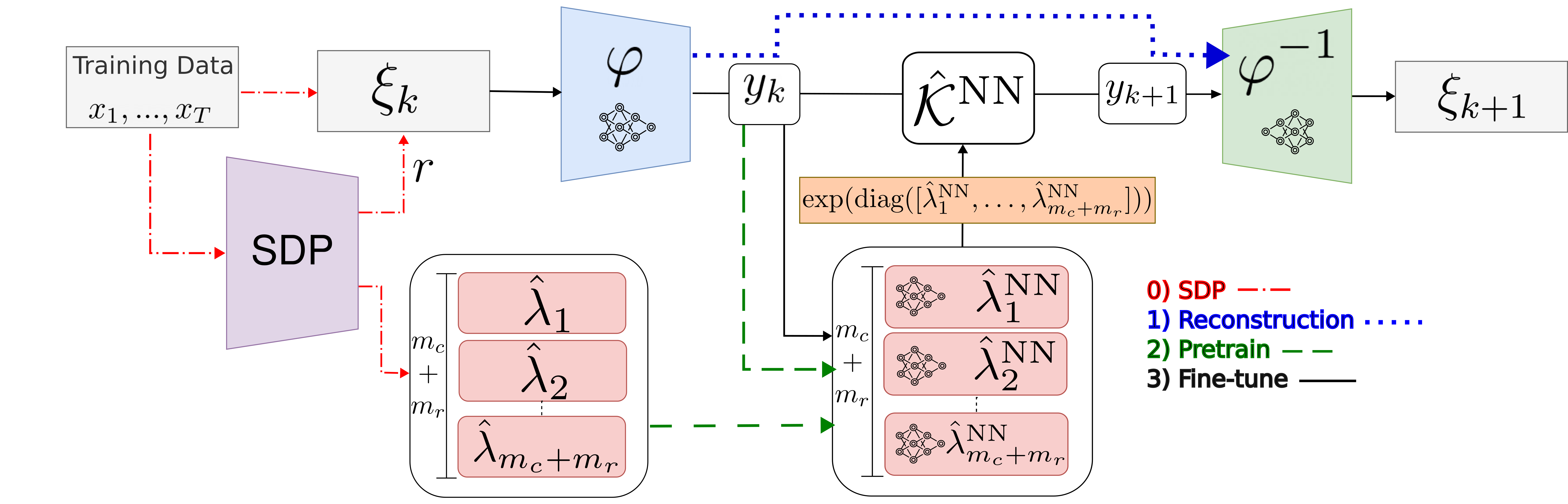} 
    \caption{Our model and training scheme. \textcolor{red}{Red} lines show the data path for solving the SDP and determining the latent space size. \textcolor{blue}{blue} lines show the reconstruction phase, which ensures recoverable states before pretraining. \textcolor{green}{Green} lines show the pretraining phase that populates the auxiliary network with the SDP-derived eigenvalues. Black lines refer to the final data path used for fine-tuning and inference.}
    \label{fig:model}
\end{figure*}

At a high level, our approach has two stages: First, by solving the SDP discussed in \cite{sznaier2021convexoptimizationapproachlearning}, we determine the dimension of the observable space, the system's memory, and an approximate Koopman operator. Then, as seen in \cref{fig:model} we use these parameters to construct the AE, avoiding prespecification. Our neural network consists of two ReLU Multilayer Perceptrons (MLPs) for the AE, and a set of auxiliary MLPs to represent the eigenvalues of the learned Koopman. For chaotic systems, we additionally consider the order found from the SDP in constructing the length of state history data used as input to the AE. 

\subsection{Koopman Extraction} \label{Koop_Extract}
Given a training dataset $\boldsymbol{X} \subset \mathbb{R}^n$, we use a subset $\boldsymbol{X}_{\text{SDP}}$ and solve the convex SDP outlined in Algorithm 1 of \cite{sznaier2021convexoptimizationapproachlearning} to determine approximate observables $\hat{\bm{y}}_n$ along with the system memory $r$. 

Since our approach learns the embedding function using an autoencoder, we simplify the SDP from~\cite{sznaier2021convexoptimizationapproachlearning} to no longer find the Loewner matrices (which defines a mapping into the observable space). Thus this modified SDP finds only a set of measurements (observables) $\hat{\bm{Y}}$ and order $r$.
Using $\hat{\bm{Y}}$ and $r$, an approximate Koopman, $\hat{\mathcal{K}}$ can be derived as outlined in \cref{time-delay-hankel} using HAVOK \cite{HAVOK17}.

\subsection{Network Setup}
To handle memory, we concatenate multiple states along a trajectory $\xi_k = \begin{bmatrix} \bm{x}_{k-r+1}^T & \cdots & \bm{x}_k^T \end{bmatrix}^T$, where $r$ is the order obtained from the SDP and $\bm{x} \in \boldsymbol{X}$.

The size of the embedded space is set by $\hat{\mathcal{K}}$. Furthermore, like in \cite{Lusch_2018}, we initialize several auxiliary networks to learn each complex and real eigenvalue of the Koopman, allowing the eigenvalues to vary. To determine $m_c$ and $m_r$, the number of complex and real eigenvalues respectively, we perform an eigen-decomposition of $\hat{\mathcal{K}}$ .

\subsection{Training Procedure}
\label{training-procedure}
To take advantage of our SDP's approximate Koopman operator, we split training into several phases:
\subsubsection{Reconstruction}
The network is first trained using only reconstruction loss \cref{eq:reconstruct_loss} to find a preliminary latent space. The autoencoder is trained simply to embed states into the latent space and decode back into the state space.

\subsubsection{Koopman Pretrain} \label{Koop_Pretrain}
Once the encoder and decoder are able to reconstruct, we freeze the autoencoder and train an auxiliary network for each of the $m_c + m_r$ eigenvalues of $\hat{\mathcal{K}}$ that maps from vectors in our latent space (embedded space) to the corresponding eigenvalue using MSE loss. We then construct a Koopman by first creating a diagonal matrix from these eigenvalues, and then taking the matrix exponential.

\subsubsection{Fine-Tuning}
Afterward the parameters of the auxiliary network are temporarily frozen while the rest of the network is trained to minimize \cref{eq:for_loss,eq:lin_loss,eq:reconstruct_loss}, aligning the latent space with the span of $\hat{\mathcal{K}}$'s eigenfunctions. This allows the network to approximate the initial SDP guess as closely as possible. Finally, we unfreeze the auxiliary network to jointly fine-tune the latent space and Koopman approximation.

\section{Experiments}\label{experiments}
By using the order and eigenvalues from the SDP in \cref{Koop_Extract}, we found that our method was able to reconstruct trajectories more accurately and in the same or less time than in \cite{Lusch_2018}. Additionally, we performed experiments to demonstrate the individual advantages of leveraging the structure found from the SDP in training an accurate neural network embedding and Koopman operator.
\subsection{Experimental Design}
\paragraph{Environments}
We demonstrate the proposed method on four dynamical systems: a Discrete Spectrum environment, a Fluid Flow on Attractor environment, a Pendulum environment, and a Lorenz attractor environment. The Discrete Spectrum serves as an example to show efficacy on discrete systems, while the Lorenz attractor serves to test efficacy on trajectories which require a long history to encode. 

The Discrete Spectrum environment has stable eigenvalues and is a common benchmark \cite{Brunton2016koop}, \cite{Tu2014}, with dynamics
 \begin{align}
     \dot{x}_1 &= \mu x_1 \\
     \dot{x}_2 &= \lambda(x_2 - x_1^2).
 \end{align}
 
 The Fluid Flow on Attractor system considers the nonlinear fluid flow past a stationary cylinder at Reynolds number 100, which is characterized by vortex shedding. This environment has been studied by Noack in \cite{Noack2003AHO}, and is a benchmark in fluid dynamics. The dynamics evolve when a low-dimensional attractor is present, producing the following model:
\begin{align}
    \dot{x}_1 &= \mu x_1 - \omega x_2 + A_1 x_3 \\
    \dot{x}_2 &= \omega x_1 + \mu x_2 + A_2 x_3 \\
    \dot{x}_3 &= -\lambda \left(x_3 - x_1^2 - x_2^2\right).
\end{align}

The nonlinear Pendulum environment exhibits a continuous eigenvalue spectrum with increasing energy, which produces the following model:
\begin{align}
    \ddot{x} = -\sin(x) \Rightarrow 
    \begin{cases}
        \dot{x}_1 = x_2 \\
        \dot{x}_2 = -\sin(x_1).
    \end{cases}
\end{align}

The Lorenz attractor is notable as a system with two attractors and chaotic solutions, where minor changes in the initial conditions can result in vastly different trajectories. As shown in \cite{sznaier2021convexoptimizationapproachlearning, HAVOK17}, solving for this system's approximate Koopman benefits from careful choice of the dynamics order. The model is a system of 3 ODEs:
\begin{align}
    \dot{x} &= \sigma(y-x) \\
    \dot{y} &= x(\rho - z) - y \\
    \dot{z} &= xy - \beta z.
\end{align}

\subsection{Experimental Setup}
For all four environments, we compare three different models by measuring the point-wise Mean Squared Error (MSE) along a set of test trajectories unseen during training. The ``Lusch" model is our re-implementation of Lusch's \cite{Lusch_2018} original work, in which we train the network from scratch with given parameters. The ``W/o Pretraining" model solves the SDP in \cref{Koop_Extract} for the dimension of the latent space and number of eigenvalues, but does not pretrain the auxiliary networks. This method's performance is meant to show the benefit of a priori knowledge of the number of eigenvalues alone to learn a given ODE. Finally, the ``W/ Pretraining" model uses the full training procedure outlined in \cref{training-procedure}. We perform two additional experiments comparing the performance of varying choices of spectral configuration and order on each system in two sweeps to reiterate the importance of the a priori choice of eigenvalues and order during learning. For the eigenvalue sweep, the SDP was not used at all, while the order sweep used the SDP and pretraining.

\subsection{Experimental Results}

\subsubsection{Resultant Trained Models}
\begin{table}[h!]
\begin{center}
\resizebox{\columnwidth}{!}{
\begin{tabular}{| c || c c c | c c c |}
\hline
Spectral Composition & \multicolumn{3}{c|}{Lusch \cite{Lusch_2018}} &\multicolumn{3}{c|}{Ours} \\
\cline{2-7}
  & \# Real & \# Complex & Order & \# Real & \# Complex & Order \\
\hhline{|=||= = =|= = = |}
Discrete Spectrum       & 2 & 0 & 1 & 2 & 0 & 1 \\
\hline
Fluid Flow On Attractor & 0 & 2 & 1 & 1 & 2 & 1 \\
\hline
Pendulum                & 0 & 2 & 1 & 0 & 2 & 1  \\
\hline
Lorenz (sweep)          & 2 & 4 & 1 & 2 & 4 & 2 \\
\hline
\end{tabular}
}
\end{center}
\caption{Spectral configuration from \cite{Lusch_2018} compared to the results of the SDP from \cref{Koop_Extract}}
\label{tab:spec_comp}
\end{table}

For the W/o Pretraining and W/ Pretraining models, the spectral configuration (number of real and complex eigenvalues) for each environment found by solving the SDP can be directly compared to that found in Lusch \cite{Lusch_2018}, except for the Lorenz attractor. To replace this baseline, a sweep of eigenvalue configurations was performed to determine the spectral configuration for the Lorenz environment. 
As seen in \cref{tab:spec_comp}, the SDP found the same spectral configuration as Lusch for both the Pendulum and Discrete Spectrum environments, meaning that the W/o Pretrain and Lusch models for those two environments returned the same model; hence, we focus our discussion on the Fluid Flow on Attractor and Lorenz environments.

\subsubsection{Prediction Performance over Test Set}
\begin{figure}[h!]
    \centering
    \includegraphics[trim=0 20 0 0 , clip,width=\linewidth]{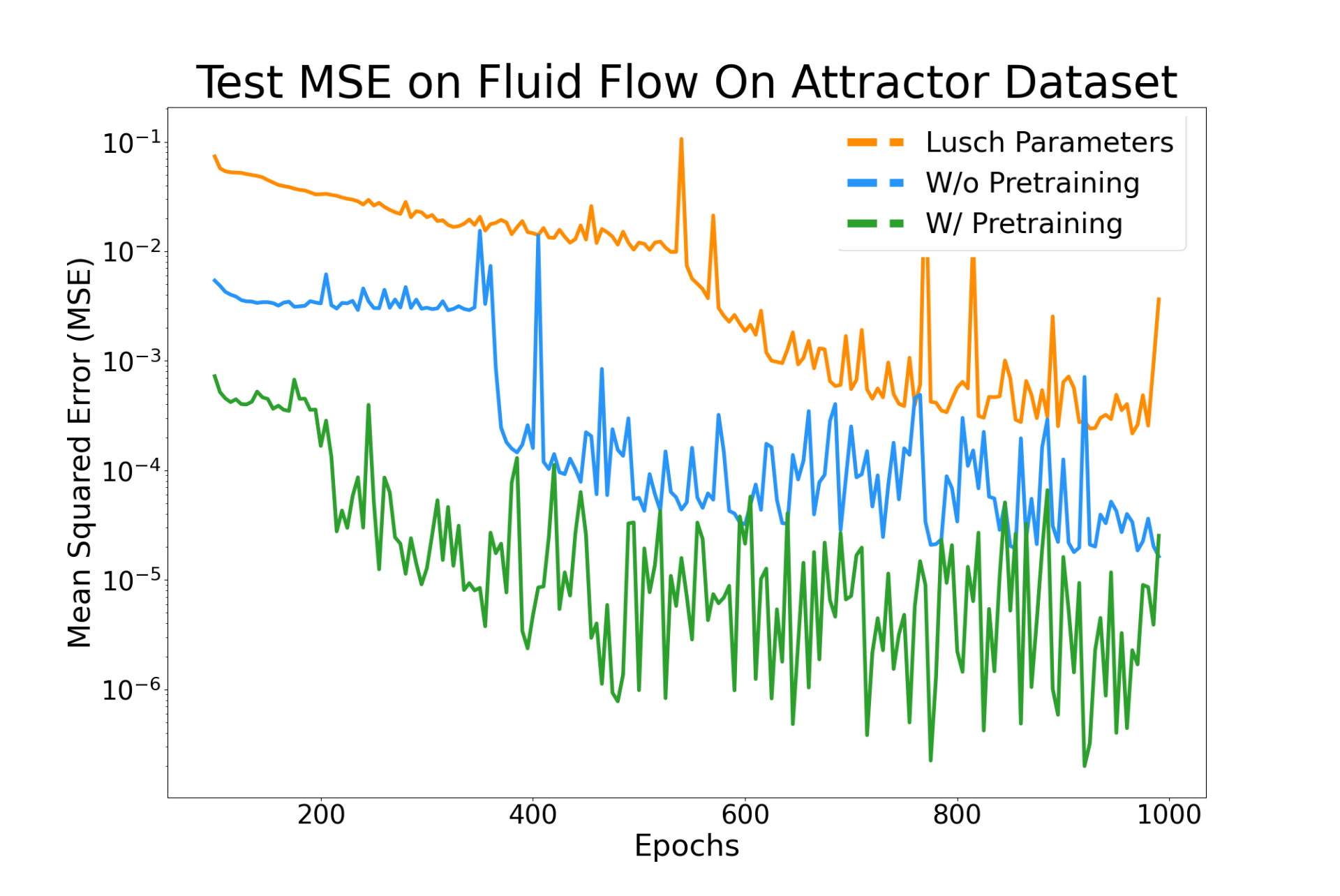}
    \includegraphics[trim=0 0 0 20 , clip, width=\linewidth]{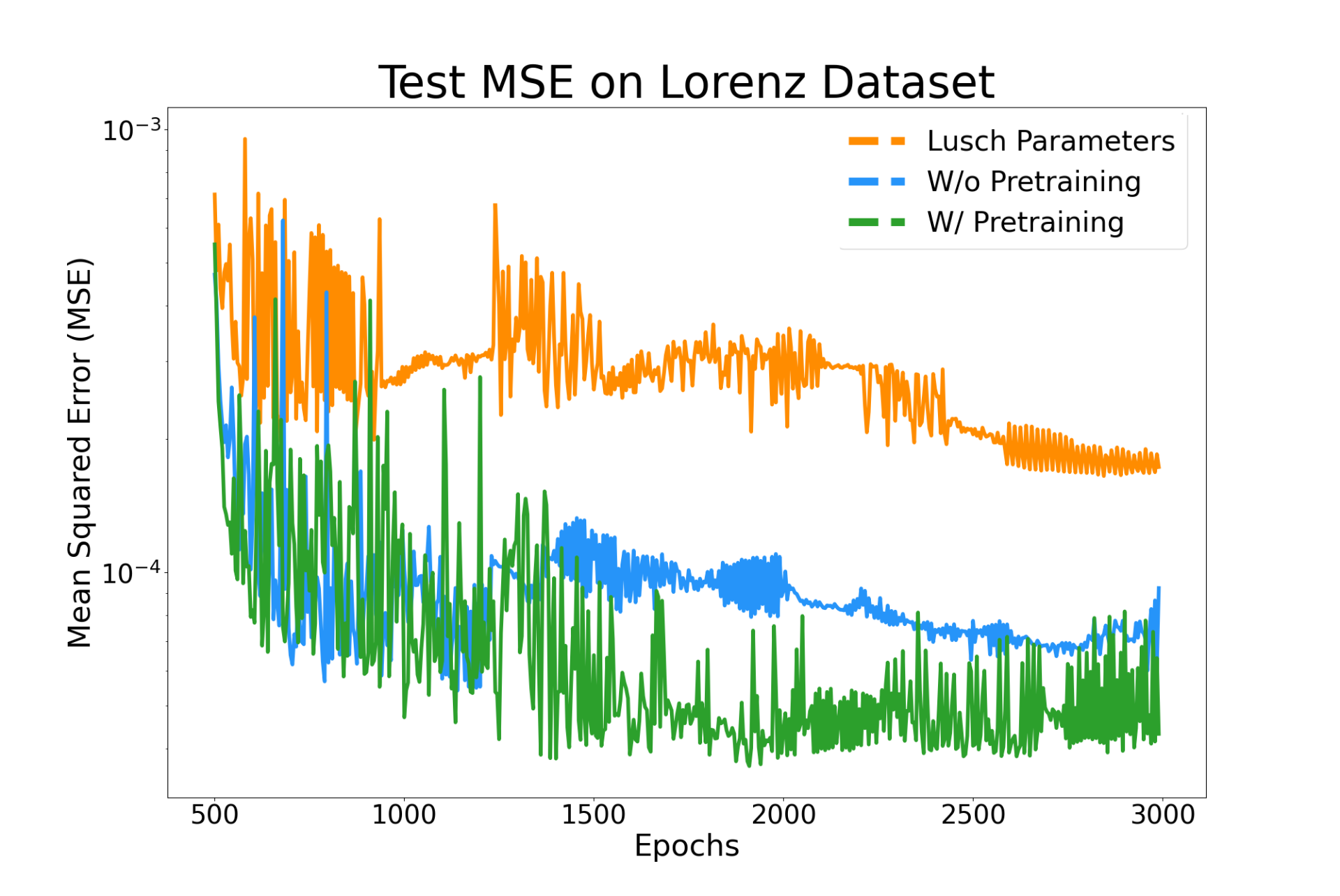}
    \caption{\textbf{MSE Curves During Training:} Our W/ Pretraining system was able to outperform both other models on 1-step prediction on the test set, while our W/o Pretraining system still out performed the Lusch model and the eigenvalue sweep.}
    \label{fig:mse_compare}
\end{figure}

\begin{figure*}[h!]
\centering
\raisebox{-5mm}{
    \begin{subfigure}[b]{0.5\linewidth}
    \centering
        \includegraphics[width=0.96\linewidth]{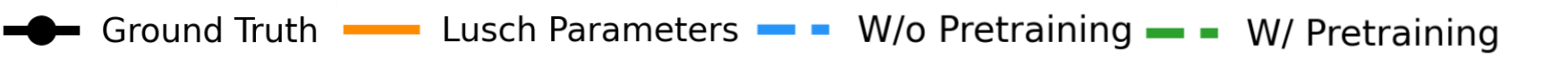}
    \end{subfigure}
    \vspace{10pt}}
\resizebox{\textwidth}{!}{
\begin{minipage}{\textwidth}
    \centering
    \begin{subfigure}[t]{0.24\textwidth}
        \centering
        \raisebox{3mm}{\includegraphics[trim=0 0 410 65, clip, width=\linewidth]{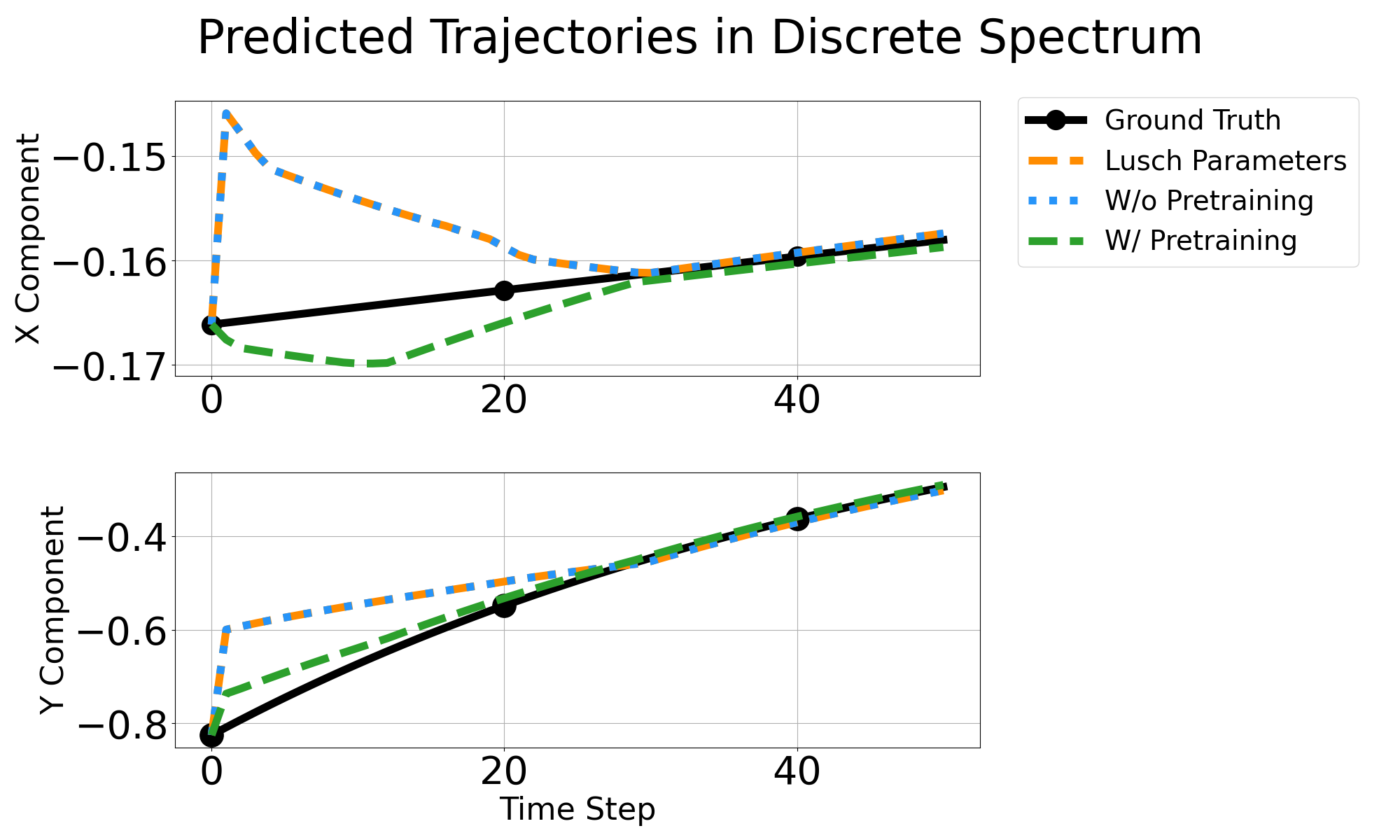}}
        \caption{Discrete Spectrum}
        \label{fig:dse_rmse_compare}
        \end{subfigure}
    \begin{subfigure}[t]{0.24\textwidth}
        \centering
        \raisebox{3mm}{\includegraphics[trim=0 0 410 65, clip, width=\linewidth]{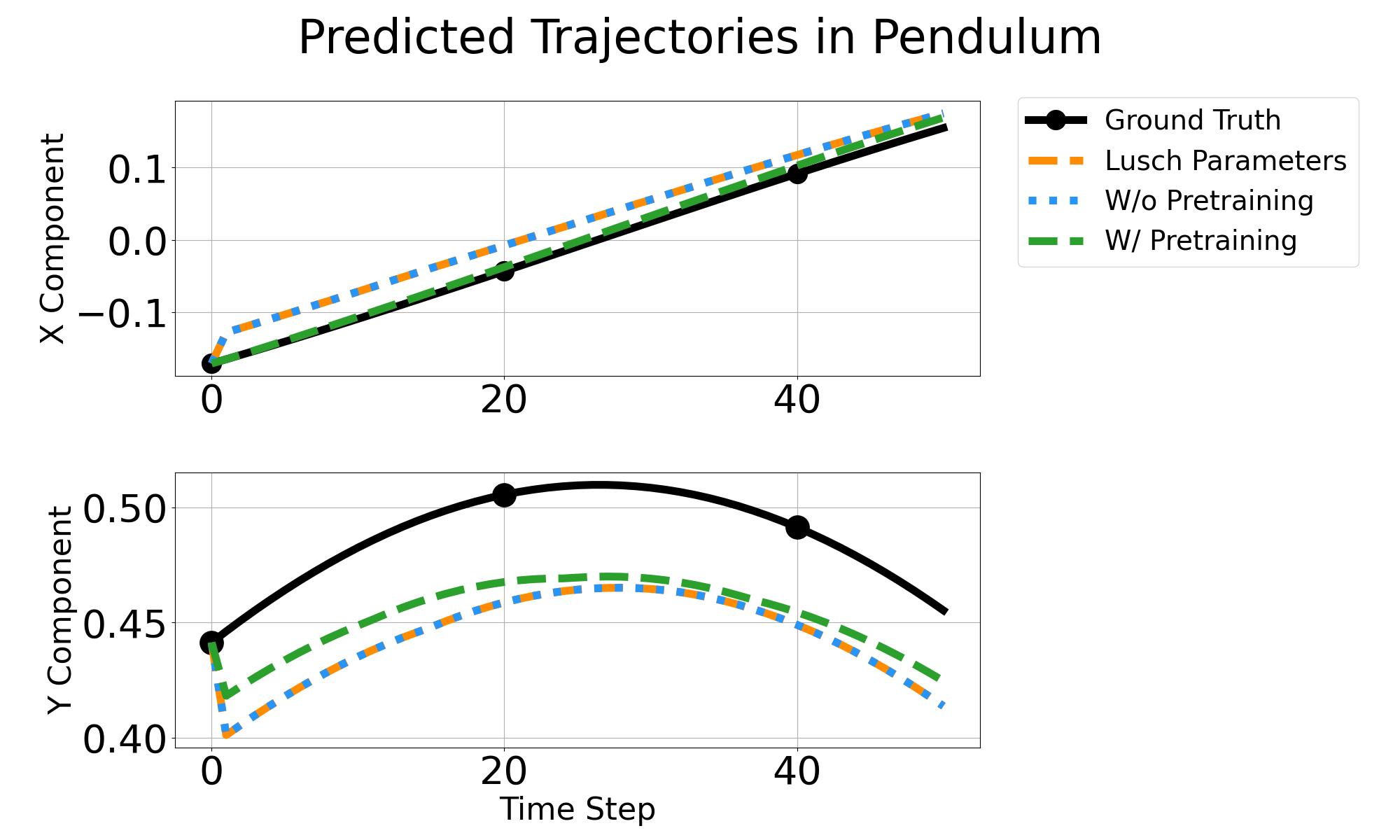}}
        \caption{Pendulum}
        \label{fig:pendulum_rmse_compare}
    \end{subfigure}
    \begin{subfigure}[t]{0.24\textwidth}
        \centering
        \includegraphics[trim=0 0 410 70, clip, width=\linewidth]{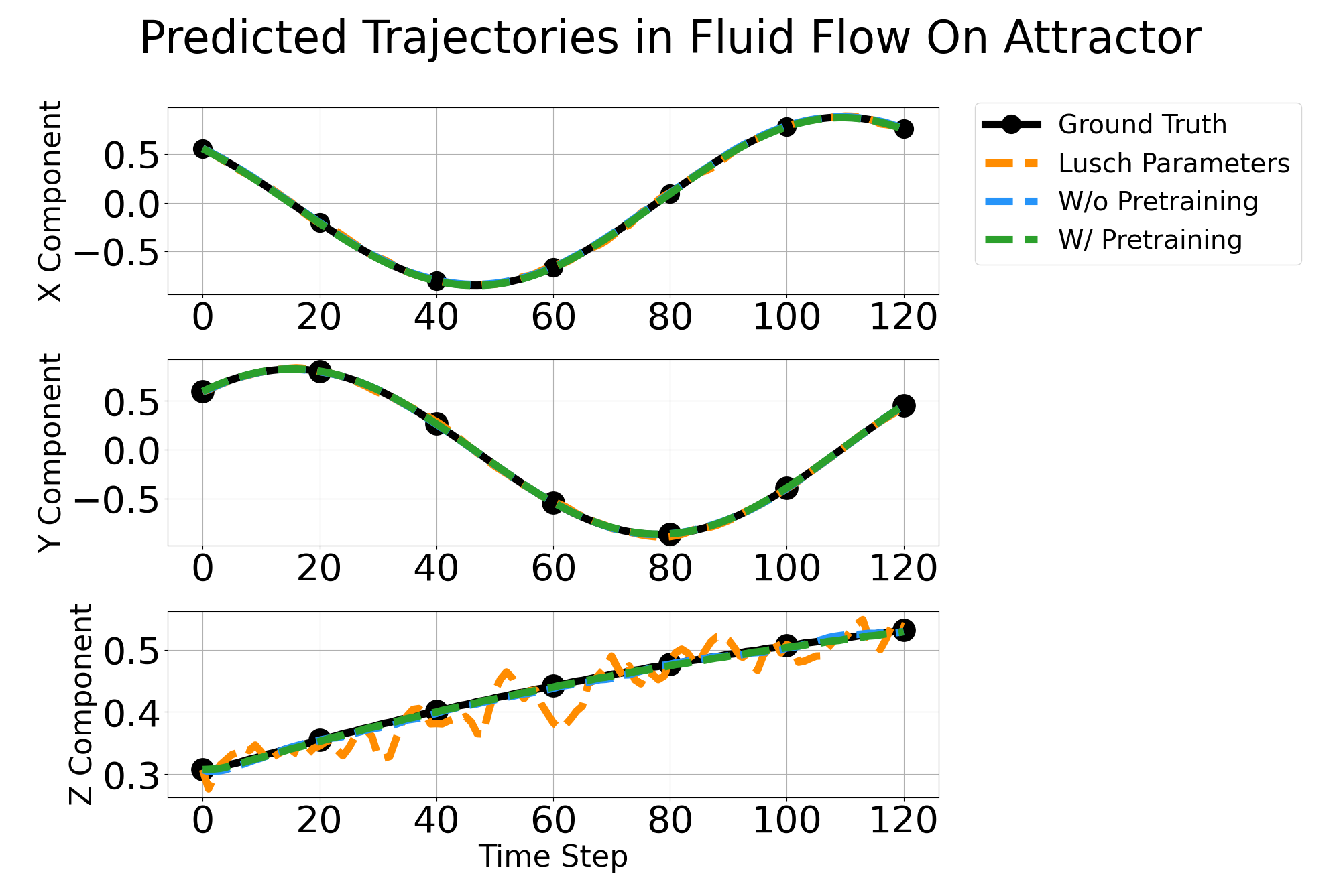}
        \caption{Fluid Flow On Attractor}
        \label{fig:ffoa_rmse_compare}
    \end{subfigure}
    \begin{subfigure}[t]{0.24\textwidth}
        \centering
        \includegraphics[trim=0 0 410 70, clip, width=\linewidth]{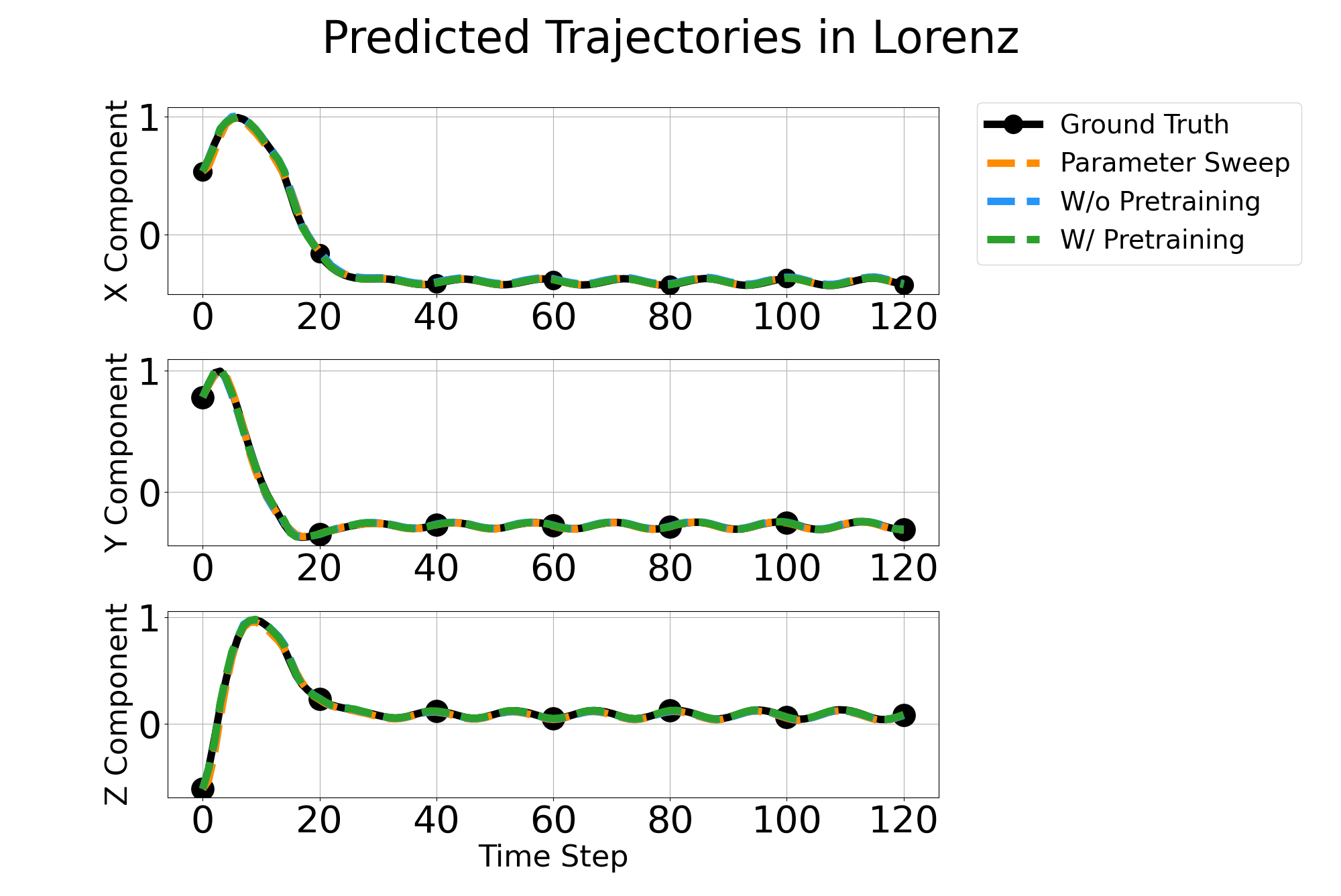}
        \caption{Lorenz}
        \label{fig:lorenz_rmse_compare}
    \end{subfigure}
    \caption{\textbf{Predicted state space trajectories for each system:} In \cref{fig:ffoa_rmse_compare}, the additional real eigenvalue found by the SDP allows for more accurate prediction of the $z$ component. Because Lorenz was not included as a baseline in \cite{Lusch_2018}, a parameter sweep was performed to determine its spectral configuration.}
    \label{fig:traj_compare}
    \end{minipage}}
    \end{figure*}

\begin{figure}[h]
    \begin{subfigure}[t]{0.24\textwidth}
        \centering
        \includegraphics[trim=0 0 20 90, clip, width=\linewidth]{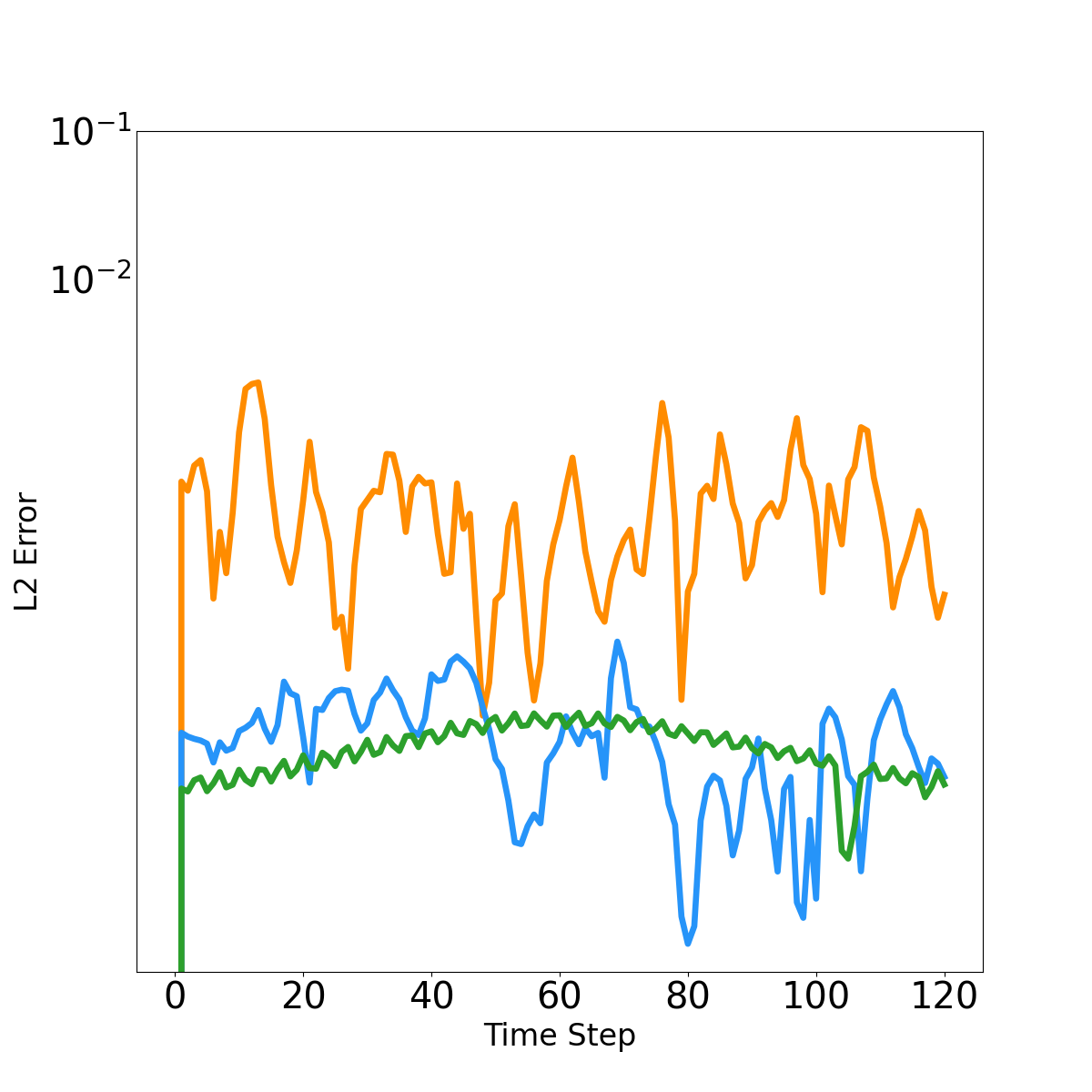}
        \caption{Fluid Flow on Attractor $L_2$ Error}
        \label{fig:quad_a}
    \end{subfigure}
    \begin{subfigure}[t]{0.24\textwidth}
        \centering
        \includegraphics[trim=0 0 30 100, clip, width=\linewidth]{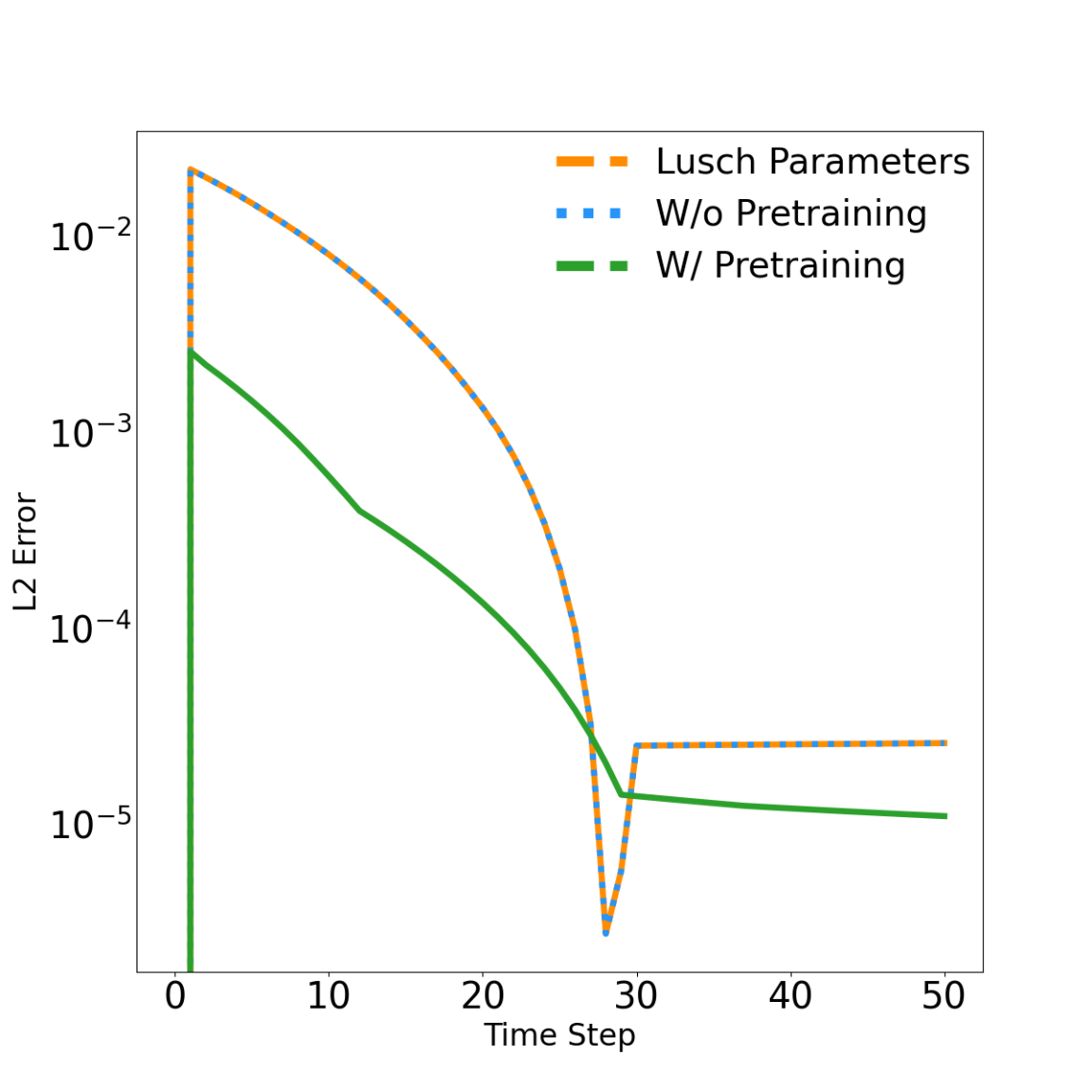}
        \caption{Discrete Spectrum $L_2$ Error}
        \label{fig:quad_b}
    \end{subfigure}
    
    \vspace{0.5cm}  
    
    \begin{subfigure}[t]{0.24\textwidth}
        \centering
        \includegraphics[trim=0 0 30 100, clip, width=\linewidth]{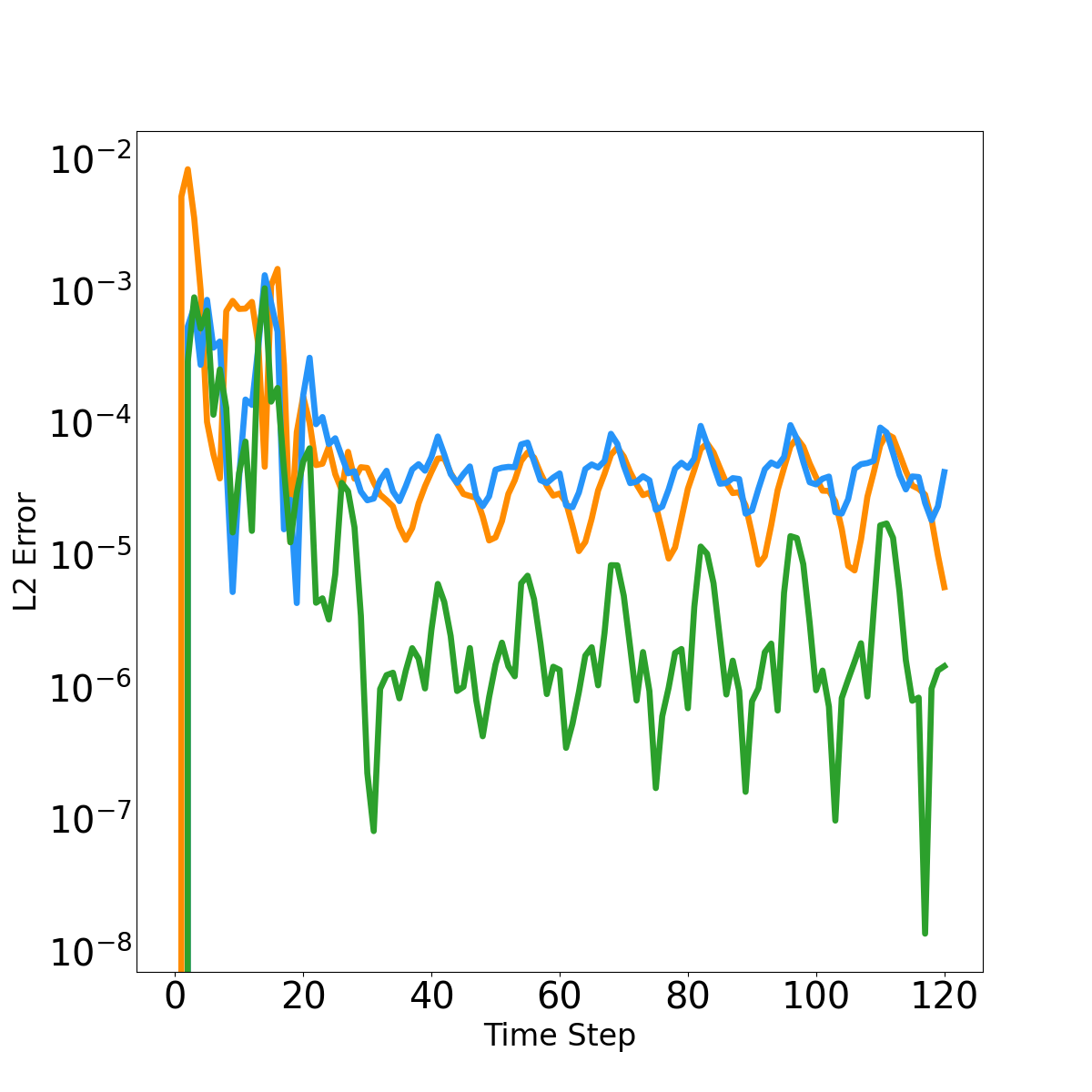}
        \caption{Lorenz $L_2$-Error}
        \label{fig:quad_c}
    \end{subfigure}
    \begin{subfigure}[t]{0.24\textwidth}
        \centering
        \includegraphics[trim=0 0 0 80, clip, width=\linewidth]{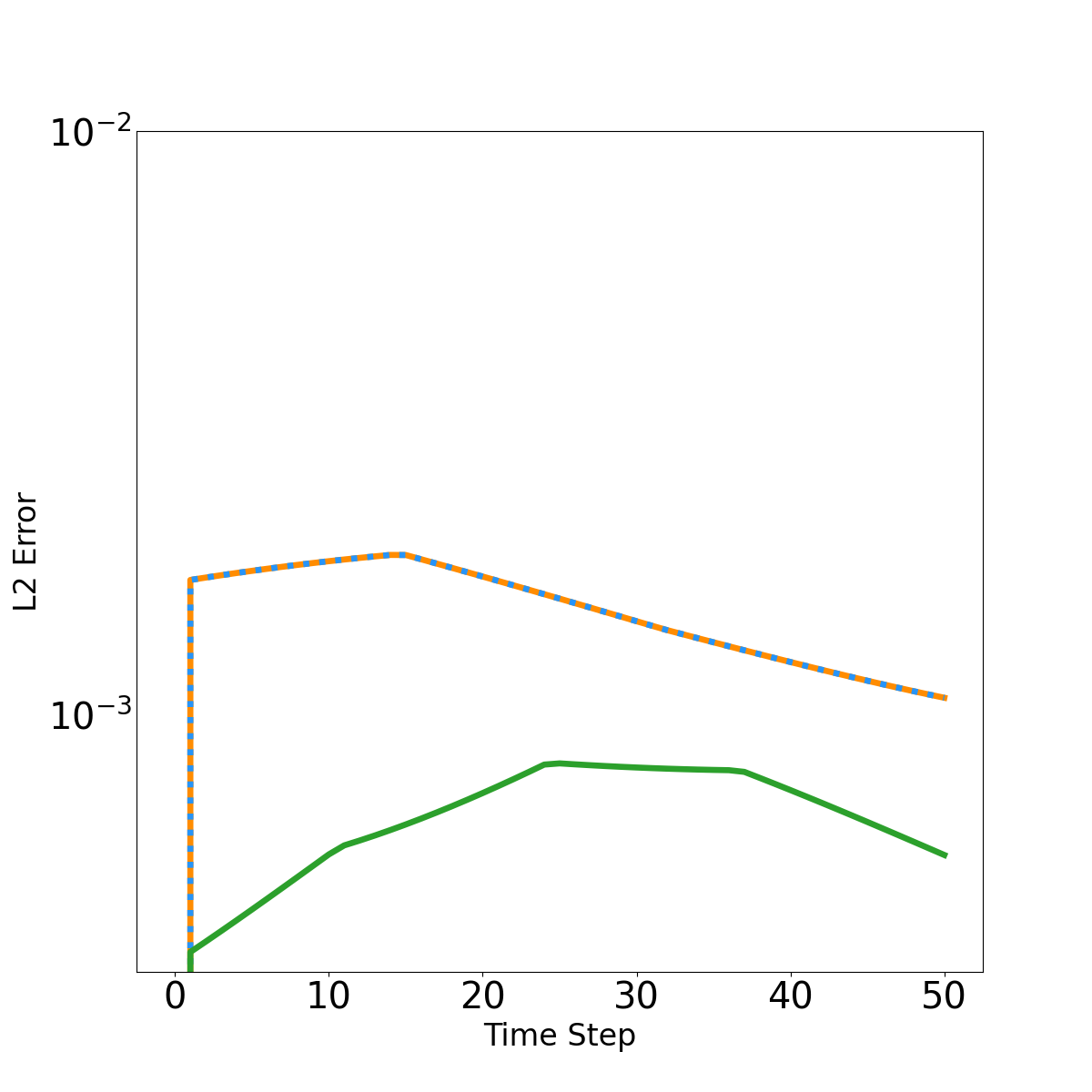}
        \caption{Pendulum $L_2$-Error}
        \label{fig:quad_d}
    \end{subfigure}
    
    \caption{\textbf{$L_2$ Error On Example Trajectory}: We display the $L_2$ error for each example trajectory vs. 1-step prediction shown in \cref{fig:traj_compare}. In general, the W/ Pretraining model has a lower $L_2$ error across all environments for most time steps}
    \label{fig:rmse_compare}
\end{figure}

\paragraph{Test MSE} \label{sec:mseexp}

We measure the MSE between the unseen test set and each model's single step predictions across each trajectory. MSE was recorded during training, and results can be seen for the Fluid Flow on Attractor and Lorenz datasets in \cref{fig:mse_compare}. In all environments, we found that our model W/ Pretraining outperforms both the Lusch and W/o Pretraining models. In the Fluid Flow on Attractor and Lorenz environments, the SDP outlined in \cref{Koop_Extract} informed the system W/o Pretraining with a better number of complex and real eigenvalues to capture the dynamics of the system than Lusch chose. When comparing the final performance of all models in \cref{tab:1stepmse}, the W / pretraining model continues to corroborate our earlier findings, sometimes outperforming the Lusch model by an order of magnitude.
As can be seen in \cref{fig:mse_compare} and \cref{fig:ffoa_rmse_compare}, the distinction between the performance of the Lusch and W/o Pretrain models on the Fluid Flow environment can be attributed to the addition of a real eigenvalue derived from the SDP. This assists the model in correcting prediction of the z-component. 

Interestingly, as seen in \cref{fig:mse_compare}, our W/o Pretrain model still had a substantially lower MSE curve compared to the Lusch model. From \cref{tab:spec_comp}, both the W/o Pretrain and Lusch models had the same spectral composition, with the only difference being the order of the system. This demonstrates that both spectral composition and order are important when finding our learned Koopman operator.



\begin{table}[h!]
\begin{center}
\resizebox{0.96\columnwidth}{!}{
\begin{tabular}{| c || c | c | c | }
\hline
 1-Step MSE & Lusch \cite{Lusch_2018} & W/o Pretraining & W/ Pretraining\\ 
 \hhline{|=||=|=|=|}
 Discrete Spectrum & 7.06e-6 & 7.06e-6 & \textbf{3.73e-6}\\
 \hline
 Fluid Flow On Attractor & 6.9e-4  & 3.11e-5  & \textbf{6.14e-6}\\
 \hline
 Pendulum & 9.93e-4 & 9.93e-4 & \textbf{3.82e-4} \\
 \hline
 Lorenz & 1.77e-4 & 7.42e-5 & \textbf{5.35e-5} \\
 \hline
\end{tabular}
}
\end{center}
\caption{\textbf{1-Step MSE of each method trained across each system:} Our W/ Pretraining model outperformed all other models in 1-step test set trajectory reconstruction.}
\label{tab:1stepmse}
\end{table}

\paragraph{Example Trajectory Prediction}
To display the performance of the three models on the test set, we have qualitatively assessed their predictive performance in \cref{fig:traj_compare} on randomly selected trajectories. In all of the simple environments, our W/ Pretraining method outperformed both Lusch and the W/o Pretraining models, corroborating our findings from above. Additionally, we include the $L_2$ Error for the same trajectories between the ground truth and single step predictions in \cref{fig:rmse_compare}. Despite the visual similarity in the accuracy of trajectory prediction shown in \cref{fig:lorenz_rmse_compare}, the $L_2$ errors of the same test in \cref{fig:quad_c} show that our W/ Pretraining method outperformed the naive eigenvalue sweep by an order of magnitude. 


\subsubsection{Training Wall-Clock Time Comparison}
To assess computational overhead for real application, we measured the wall-clock time required to fully train the models. Both SDP-informed systems converged more quickly than the Lusch model (Fluid Flow environment) and eigenvalue sweep (Lorenz environment). When SDP runtime is factored in, the chaotic environment Lorenz took the same amount of time to run as our W/ Pretraining model, however performed much worse in prediction accuracy. The W/o Pretraining model which uses the SDP-informed structure only was the fastest, but converged early and did not perform as well on prediction as the W/ Pretraining model. In the Fluid Flow On Attractor environment, W/ Pretraining converged significantly faster than both other methods.

\begin{table}[h!]
\centering
\raisebox{-13mm}{
\resizebox{0.96\columnwidth}{!}{
\begin{tabular}{| c || c | c | c| }
\hline
 Lorenz Runtime (min) & SDP & Training & Total Time \\ 
 \hhline{|=||=|=|=|}
 Lusch \cite{Lusch_2018} & - & 73 & 73\\
 \hline
 W/o Pretraining &  6 & 42 & \textbf{48}\\
 \hline
 W/ Pretraining & 6 & 67 & 73\\
 \hline
\end{tabular}
}
}
\caption{\textbf{Runtime Comparison:} The run-times of the system chosen by eigenvalue sweep and W/ Pretraining method are equivalent for this complex environment. Our W/o Pretraining method performed well in wall-clock time, however converged at a worse performance than W/ Pretraining.}
\label{tab:lorenz}
\end{table}

\begin{table}[h]
\centering
\resizebox{0.96\columnwidth}{!}{
\begin{tabular}{| c || c | c | c| }
\hline
 Fluid Flow Runtime (min) & SDP & Training & Total Time \\ 
 \hhline{|=||=|=|=|}
 Lusch \cite{Lusch_2018} & - & 33 & 33 \\
 \hline
 W/o Pretraining & 5  & 13 & 18 \\
 \hline
 W/ Pretraining & 5 & 7 & \textbf{12 }\\
 \hline
\end{tabular}
}
\caption{\textbf{Runtime Comparison:} In the Fluid Flow environment, our model W/ Pretraining converged the quickest, with Lusch never having converged in the 1000 epochs allotted and taking the maximum time.}
\label{tab:fluidflow}
\end{table}

\subsubsection{Eigenvalue Sweep}
\begin{figure}[h!]
    \centering
    \raisebox{-30mm}{
    \resizebox{0.96\columnwidth}{!}{
    \includegraphics[trim=0 0 0 8, clip, width=\linewidth]{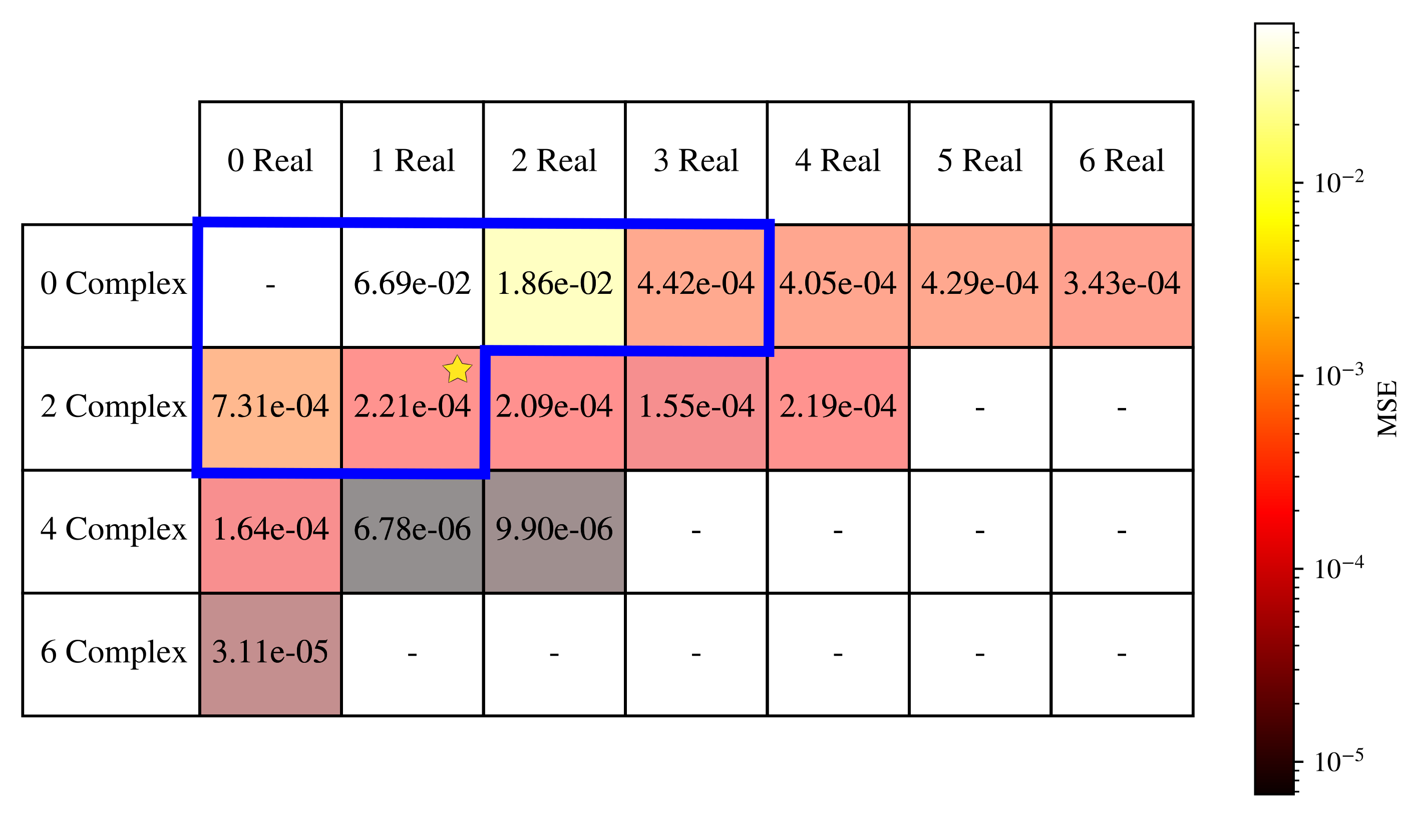}}}
    \caption{\textbf{Final Test MSE for Fluid Flow on Attractor (order 1) for varying number of eigenvalues:} \textcolor{blue}{Blue} outline marks models with an embedded dimension of 3 or less. The star indicates the SDP-derived spectral configuration, which is the best for total number of eigenvalues up to 3, with diminishing returns until dimension 5.}
    \label{fig:ffoa_eigsweep}
\end{figure}

\begin{figure}[h!]
    \centering
    \raisebox{-10mm}{
    \resizebox{0.98\columnwidth}{!}{
    \includegraphics[width=\linewidth]{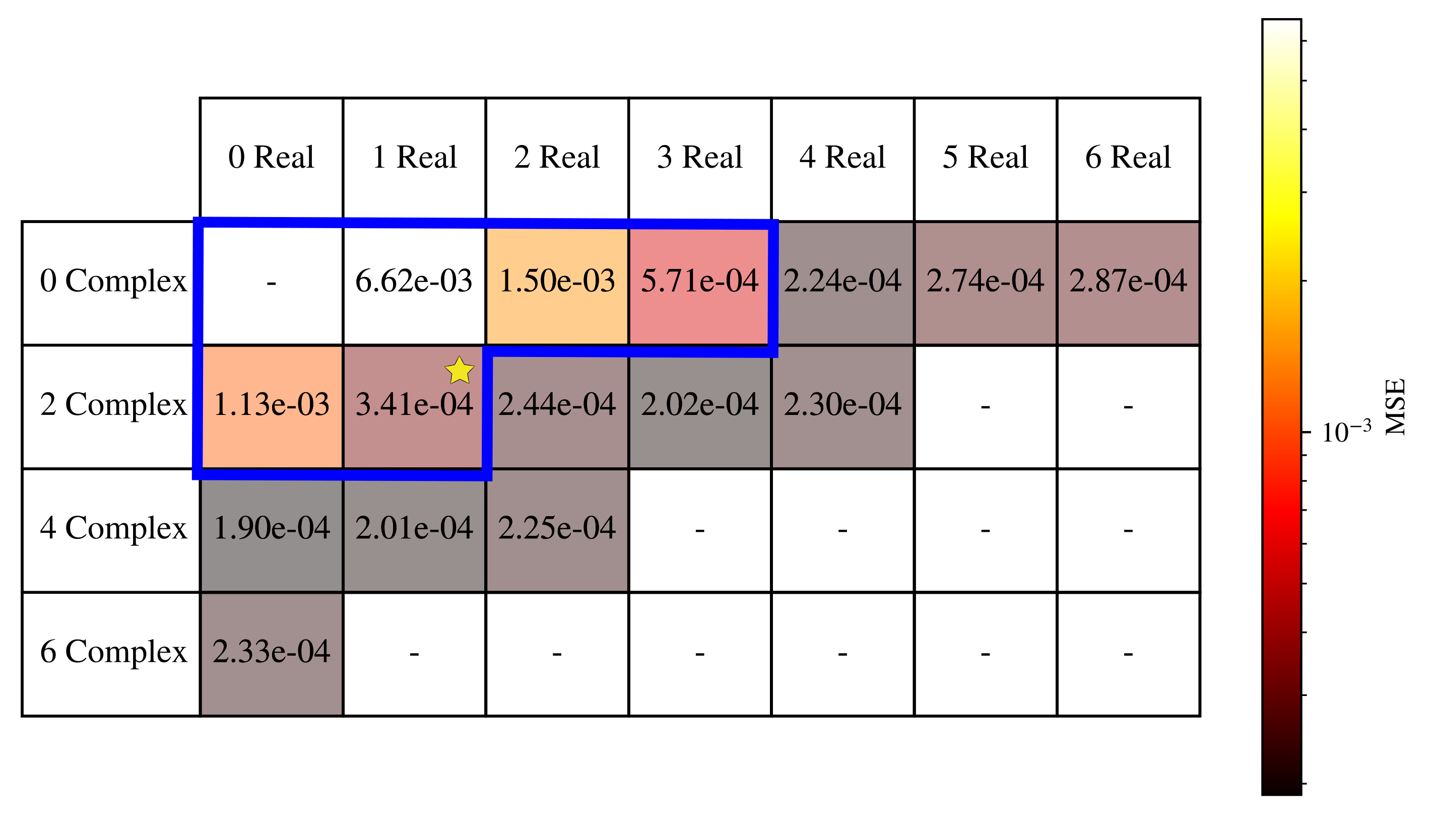}}}
    \caption{\textbf{Final Test MSE for Lorenz oscillator (order 1) for varying number of eigenvalues:} 
    \textcolor{blue}{Blue} outline marks models with an embedded dimension of 3 or less. The star indicates the SDP-derived spectral configuration, which is the best for total number of eigenvalues up to 3, with diminishing returns for higher dimensions.
    }
    \label{fig:lorenz-eigsweep}
\end{figure}

To investigate the importance of choosing the correct number of complex and real eigenvalues before training, we evaluated all possible configurations of real and complex eigenvalues up to 6 eigenvalues total; as complex eigenvalues come in pairs, this resulted in 15 different trials. As seen in \cref{fig:ffoa_eigsweep} and \cref{fig:lorenz-eigsweep}, for configurations with up to 3 dimensions, both the Fluid Flow on Attractor and Lorenz systems performed best with 2 complex and 1 real eigenvalues. These results match our SDP-derived spectral configuration. Although several configurations did result in lower MSE, configurations that resulted in an order of magnitude improvement needed at least two more eigenvalues, almost doubling the number of auxiliary networks needed to train. Furthermore, the model with our SDP-derived spectral configuration outperformed models with 4, 5, and 6 real eigenvalues. This demonstrates that even with more eigenvalues, without using a suitable spectral configuration, prediction may not necessarily improve.

\subsubsection{Order Sweep}
\begin{figure}[H]
    \centering
    \raisebox{-10mm}{
    \resizebox{0.96\columnwidth}{!}{
    \includegraphics[trim=0 0 0 54,clip,width=\linewidth]{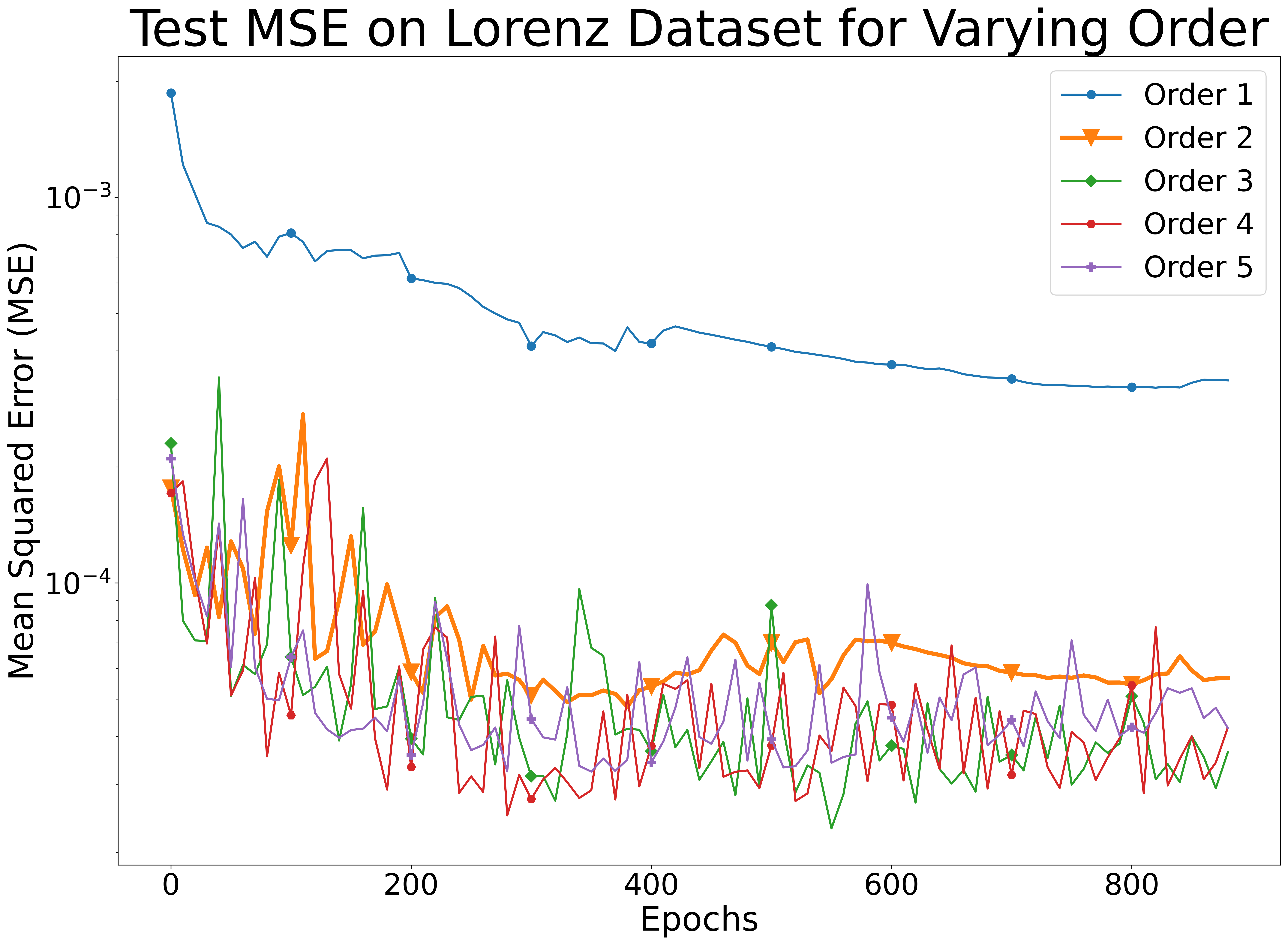}}}
    \caption{\textbf{Test MSE for Lorenz of varying order:} The \textcolor{orange}{orange} line shows the SDP-derived order, yielding an order-of-magnitude improvement over the single-order system.} 
    \label{fig:lorenz_ordersweep}
\end{figure}

Similarly to the eigenvalue sweep, we trained each of the models on the chaotic Lorenz oscillator system, but with prespecifying the order of the dynamics for the found Koopman. As can be seen in \cref{fig:lorenz_ordersweep}, there is an order of magnitude difference in MSE between the order 1 dynamics model and all other models. There is a smaller difference between order 2 and order 3-5, with all other models having very similar MSE curves. Again, this shows the importance of accurately specifying the underlying order. With too small of an order, the dynamics may have significantly larger loss, and with too large of an order, there are diminishing returns and the network must become bigger to handle larger input, latent, and output spaces. The SDP from \cref{sec:mseexp} found a model using order 2 dynamics for the W/ Pretraining and W/o Pretrain experiments, which benefits from the majority of the performance benefit over order 1, but keeps the network small. 


\section{Conclusion}
This work proposes a hybrid approach to finding a Koopman operator and embedding function by combining an SDP method, which requires no a priori knowledge of the embedding, and a learning approach which leverages the advantages of training on large amounts of data. In this process, we avoid both the prespecification requirement of the learning approach and excessive scaling factors of SDP computation. We found that in all environments originally assessed by \cite{Lusch_2018} as well as a characteristically chaotic environment, our hybrid approach outperformed both the baseline and an SDP-assisted model given the correct embedding structure. Our results additionally show that a priori knowledge of the best complex and real eigenvalue frequency is imperative for accurate results on trajectory reconstruction and prediction. Potential future extensions to our work include using new state of the art neural network architectures such as \cite{gu2023mamba} and \cite{gu2020hippo} as a replacement for our MLP architecture to handle memory. Another interesting direction is in trying to explicitly learn discontinuous linear embeddings to globally handle nonlinear systems with multiple limit sets \cite{LIU202360}. Our results provide support that for dynamics model approximation, by combining the structure found from optimization methods with the large data scalability of learning based methods, one can benefit from the advantages of both.

\bibliographystyle{IEEEtran}
\bibliography{bib}

\begin{thebibliography}{10}
\providecommand{\url}[1]{#1}
\csname url@samestyle\endcsname
\providecommand{\newblock}{\relax}
\providecommand{\bibinfo}[2]{#2}
\providecommand{\BIBentrySTDinterwordspacing}{\spaceskip=0pt\relax}
\providecommand{\BIBentryALTinterwordstretchfactor}{4}
\providecommand{\BIBentryALTinterwordspacing}{\spaceskip=\fontdimen2\font plus
\BIBentryALTinterwordstretchfactor\fontdimen3\font minus \fontdimen4\font\relax}
\providecommand{\BIBforeignlanguage}[2]{{%
\expandafter\ifx\csname l@#1\endcsname\relax
\typeout{** WARNING: IEEEtran.bst: No hyphenation pattern has been}%
\typeout{** loaded for the language `#1'. Using the pattern for}%
\typeout{** the default language instead.}%
\else
\language=\csname l@#1\endcsname
\fi
#2}}
\providecommand{\BIBdecl}{\relax}
\BIBdecl

\bibitem{doi:10.1073/pnas.17.5.315}
\BIBentryALTinterwordspacing
B.~O. Koopman, ``Hamiltonian systems and transformation in hilbert space,'' \emph{Proceedings of the National Academy of Sciences}, vol.~17, no.~5, pp. 315--318, 1931. [Online]. Available: \url{https://www.pnas.org/doi/abs/10.1073/pnas.17.5.315}
\BIBentrySTDinterwordspacing

\bibitem{SCHMID_2010}
P.~J. SCHMID, ``Dynamic mode decomposition of numerical and experimental data,'' \emph{Journal of Fluid Mechanics}, vol. 656, p. 5–28, 2010.

\bibitem{ROWLEY_MEZIĆ_BAGHERI_SCHLATTER_HENNINGSON_2009}
C.~W. ROWLEY, I.~MEZIĆ, S.~BAGHERI, P.~SCHLATTER, and D.~S. HENNINGSON, ``Spectral analysis of nonlinear flows,'' \emph{Journal of Fluid Mechanics}, vol. 641, p. 115–127, 2009.

\bibitem{jose_nathan_kutz_brunton_brunton_proctor_2017}
J.~N. Kutz, S.~L. Brunton, B.~W. Brunton, and J.~L. Proctor, \emph{Dynamic mode decomposition : data-driven modeling of complex systems}, 9781611974492, Ed.\hskip 1em plus 0.5em minus 0.4em\relax Society For Industrial And Applied Mathematics, 2017.

\bibitem{Williams_2015}
\BIBentryALTinterwordspacing
M.~O. Williams, I.~G. Kevrekidis, and C.~W. Rowley, ``A data–driven approximation of the koopman operator: Extending dynamic mode decomposition,'' \emph{Journal of Nonlinear Science}, vol.~25, no.~6, p. 1307–1346, Jun. 2015. [Online]. Available: \url{http://dx.doi.org/10.1007/s00332-015-9258-5}
\BIBentrySTDinterwordspacing

\bibitem{Lusch_2018}
\BIBentryALTinterwordspacing
B.~Lusch, J.~N. Kutz, and S.~L. Brunton, ``Deep learning for universal linear embeddings of nonlinear dynamics,'' \emph{Nature Communications}, vol.~9, no.~1, Nov. 2018. [Online]. Available: \url{http://dx.doi.org/10.1038/s41467-018-07210-0}
\BIBentrySTDinterwordspacing

\bibitem{doi:10.1073/pnas.1906995116}
\BIBentryALTinterwordspacing
K.~Champion, B.~Lusch, J.~N. Kutz, and S.~L. Brunton, ``Data-driven discovery of coordinates and governing equations,'' \emph{Proceedings of the National Academy of Sciences}, vol. 116, no.~45, pp. 22\,445--22\,451, 2019. [Online]. Available: \url{https://www.pnas.org/doi/abs/10.1073/pnas.1906995116}
\BIBentrySTDinterwordspacing

\bibitem{8815339}
E.~Yeung, S.~Kundu, and N.~Hodas, ``Learning deep neural network representations for koopman operators of nonlinear dynamical systems,'' in \emph{2019 American Control Conference (ACC)}, 2019, pp. 4832--4839.

\bibitem{doi:10.1137/18M1177846}
\BIBentryALTinterwordspacing
S.~E. Otto and C.~W. Rowley, ``Linearly recurrent autoencoder networks for learning dynamics,'' \emph{SIAM Journal on Applied Dynamical Systems}, vol.~18, no.~1, pp. 558--593, 2019. [Online]. Available: \url{https://doi.org/10.1137/18M1177846}
\BIBentrySTDinterwordspacing

\bibitem{potter2022dynamics}
M.~Potter, I.~Y. Potter, O.~I. Camps, and M.~Sznaier, ``Dynamics-aware representation learning via multivariate time series transformers.'' in \emph{ESANN}, 2022.

\bibitem{sznaier2021convexoptimizationapproachlearning}
\BIBentryALTinterwordspacing
M.~Sznaier, ``A convex optimization approach to learning koopman operators,'' 2021. [Online]. Available: \url{https://arxiv.org/abs/2102.03934}
\BIBentrySTDinterwordspacing

\bibitem{BEVANDA2021197}
\BIBentryALTinterwordspacing
P.~Bevanda, S.~Sosnowski, and S.~Hirche, ``Koopman operator dynamical models: Learning, analysis and control,'' \emph{Annual Reviews in Control}, vol.~52, pp. 197--212, 2021. [Online]. Available: \url{https://www.sciencedirect.com/science/article/pii/S1367578821000729}
\BIBentrySTDinterwordspacing

\bibitem{LIU202360}
\BIBentryALTinterwordspacing
Z.~Liu, N.~Ozay, and E.~D. Sontag, ``On the non-existence of immersions for systems with multiple omega-limit sets,'' \emph{IFAC-PapersOnLine}, vol.~56, no.~2, pp. 60--64, 2023, 22nd IFAC World Congress. [Online]. Available: \url{https://www.sciencedirect.com/science/article/pii/S2405896323018165}
\BIBentrySTDinterwordspacing

\bibitem{HAVOK17}
\BIBentryALTinterwordspacing
S.~Brunton, B.~Brunton, and J.~e.~a. Proctor, ``Steven l. brunton and bingni w. brunton and joshua l. proctor and eurika kaiser and j. nathan kutz,'' \emph{Nat Commun 8, 19}, 2017. [Online]. Available: \url{https://doi.org/10.1038/s41467-017-00030-8}
\BIBentrySTDinterwordspacing

\bibitem{ionita2013lagrange}
\BIBentryALTinterwordspacing
A.~Ionita, ``Lagrange rational interpolation and its applications to approximation of large-scale dynamical systems,'' Dissertation, Rice University, 2013. [Online]. Available: \url{https://hdl.handle.net/1911/77180}
\BIBentrySTDinterwordspacing

\bibitem{Brunton2016koop}
S.~L. Brunton, B.~W. Brunton, J.~L. Proctor, and J.~N. Kutz, ``Koopman invariant subspaces and finite linear representations of nonlinear dynamical systems for control,'' \emph{PLoS ONE}, vol.~11, no.~3, p. e0150171, 2016.

\bibitem{Tu2014}
J.~H. Tu, C.~W. Rowley, D.~M. Luchtenburg, S.~L. Brunton, and J.~N. Kutz, ``On dynamic mode decomposition: theory and applications,'' \emph{Journal of Computational Dynamics}, vol.~1, no.~2, pp. 391--421, 2014.

\bibitem{Noack2003AHO}
\BIBentryALTinterwordspacing
B.~R. Noack, K.~Afanasiev, M.~Morzynski, G.~Tadmor, and F.~Thiele, ``A hierarchy of low-dimensional models for the transient and post-transient cylinder wake,'' \emph{Journal of Fluid Mechanics}, vol. 497, pp. 335 -- 363, 2003. [Online]. Available: \url{https://api.semanticscholar.org/CorpusID:55584094}
\BIBentrySTDinterwordspacing

\bibitem{gu2023mamba}
A.~Gu and T.~Dao, ``Mamba: Linear-time sequence modeling with selective state spaces,'' \emph{arXiv preprint arXiv:2312.00752}, 2023.

\bibitem{gu2020hippo}
A.~Gu, T.~Dao, S.~Ermon, A.~Rudra, and C.~R{\'e}, ``Hippo: Recurrent memory with optimal polynomial projections,'' \emph{Advances in neural information processing systems}, vol.~33, pp. 1474--1487, 2020.

\end{thebibliography}

\end{document}